\documentclass[epj,draft]{svjour}
\usepackage{latexsym}
\usepackage{graphics}
\begin{document}
\def\be{\begin{equation}}
\def\ee{\end{equation}}
\def\bea{\begin{eqnarray}}
\def\eea{\end{eqnarray}}
\title{Statistical thermodynamics of economic systems}
\author{Hernando Quevedo \inst{1} \and Mar\'\i a N. Quevedo\inst{2}
}                     
%
%
\institute{
Instituto de Ciencias
Nucleares, Universidad Nacional Aut\'onoma de M\'exico,
 AP 70543, M\'exico, DF 04510, Mexico\\
Dipartimento di Fisica and ICRA, Universit\`a di Roma "La
Sapienza",  I-00185 Roma, Italy
\and 
Departamento de Matem\'aticas, 
Universidad Militar Nueva Granada, 
Cra. 11 No. 101-80, Bogot\'a D.E., Colombia
}
\date{Received: \today / Revised version: date}
%
\abstract{
We formulate the thermodynamics of economic systems in terms of an
arbitrary probability distribution for a conserved economic quantity.
As in statistical physics, thermodynamic macroeconomic variables emerge as the mean value 
of microeconomic variables, and their determination is reduced 
to the computation of the partition function, starting from an arbitrary function.
Explicit hypothetical examples
are given which include linear and nonlinear economic systems, as well as multiplicative systems
such as those dominated by a Pareto law distribution. It is shown that the macroeconomic variables
can be drastically changed by choosing the microeconomic variables in an appropriate manner. 
We propose to use the  formalism of phase transitions to study severe changes of macroeconomic variables.
} 
\maketitle

\section{Introduction}
\label{sec:int}

Thermodynamics is a phenomenological science that derives its concepts  directly from observation and experiment. 
The laws of thermodynamics can be considered as axioms of a mathematical model, and the fact that they are based upon commonplace 
observations makes them tremendously powerful and generally valid. In particular, the interest of applying thermodynamics in a systematic manner
to describe the behavior of economic and financial systems has a long history \cite{gui32}. 
One of the difficulties of this approach is that it is necessary to 
identify {\it a priori} the economic variables that can be identified as thermodynamic variables, satisfying the laws of thermodynamics. 
The results of this identification are very often controversial. For instance, whereas in some studies money is considered as a well-defined 
thermodynamic variable, other analysis suggest that money is a completely irrelevant economic variable that, consequently, 
should not be used in any thermodynamic approach to economy \cite{lux}. 
Even the basic thermodynamic assumption that an economic system  be in 
equilibrium has been the subject of numerous discussions. 
The concept of economic entropy presents also certain difficulties and different 
definitions can be formulated \cite{mont83,smith05}. Formal mappings between thermodynamic and economic variables can be 
formulated \cite{saslow99} which, however, leave  the notion of entropy unclear and the range of models in which it holds undefined.

On the other hand, econophysics is a relatively new branch of physics \cite{stanley96} in which several methods of applied theory of probabilities,
that have been used with excellent results in statistical physics, are implemented to solve problems in economics and finance. In our opinion,
one of the most prominent founds is that certain economic variables can be considered as conserved and their distribution among the agents of 
an economic system are described by simple probability densities. In fact, it is currently well established that  wealth 
distribution in many societies presents essentially two phases
\cite{yakov00,chak00,boumez00,yakov01a,yakov01b,yakov03,silva05,chat05,chak06}. This means that the society
can be differentiated in two disjoint populations with two different probability distributions. 
Various analysis of real economic data from several countries
 have shown that one phase possesses a Boltzmann-Gibbs (exponential) probability distribution
that involves about 95\% of individuals, mostly those with medium and low wealths, 
whereas  the second phase, consisting of about 5\% of individuals
with highest wealths, shows a Pareto (power law) probability distribution.  
Similar results are found for the distribution of money and income. 

In this work, we propose to use the standard formulation of statistical thermodynamics in order to relate in a systematic and rigorous way
the thermodynamic approach with the statistical properties of complex economic systems found in econophysics. 
We obtain as a result that the entire properties of economic systems can, in principle, be formulated and considered in the definition of a  
partition function from which all the  thermodynamic properties of the system can be derived. It turns out that 
in certain systems the macroeconomic variables can be manipulated in an arbitrary way by changing the values of specific  
microeconomic variables. Moreover, the drastic changes that occur at the level of the macroeconomic variables can 
be investigated by using the formalism of phase transitions. 
 This paper is organized as
follows. In Sec. \ref{sec:td} we formulate the fundamentals of statistical thermodynamics in the case of an arbitrary 
conserved economic variable. In Sec. \ref{sec:lin} we analyze the simplest case of an economic system for which the conserved
variable depends linearly on the microeconomic parameters. More general situations are analyzed in Secs. \ref{sec:nonl} and \ref{sec:mul}
where quadratic and multiplicative dependences of abstract parameters are considered. The partition function is analyzed in all the 
cases and possible interpretations are presented. In Sec. \ref{sec:ptr} we propose to associate drastic modifications of macroeconomic 
variables with the occurrence of phase transitions. Finally, Sec. \ref{sec:con} contains  a summary and a discussion of our results.

\section{Statistical thermodynamics} 
\label{sec:td}

Consider a hypothetical economic system in equilibrium for which a quantity, say $M$, is conserved. There is a reasonable
number of arguments \cite{yakov07} which show that certain current economies can be considered as 
systems in equilibrium and some quantities, like the total 
amount of money in the system, are conserved during certain periods of time. For the sake of concreteness we will consider in this work
that $M$ is the conserved money, although our approach can be applied to any conserved quantity.
Suppose that the system is composed of $N$ agents which compete 
to acquire a participation $m$ of $M$.  In real economic systems, the total number of agents is such a large number that for most applications
the limit $N\rightarrow\infty$ is appropriate. In a closed economic system, the equilibrium probability distribution (density function) 
of $m$ is given by the Boltzmann-Gibbs distribution $\rho(m) \propto e^{-m/T}$, where $T$  is an effective temperature equal to
the average amount of money per agent. 
The amount of money $m$ that an agent can earn depends on several additional parameters 
$\lambda_1,\lambda_2,...,\lambda_l$,
which we call {\it microeconomic} parameters. Since the density function can be normalized to 1, we obtain \cite{huang}
\be
\rho(\bar\lambda) = \frac{e^{-m(\bar\lambda)/T}} {Q(T,\bar x)}\ , \quad
Q(T,\bar x) = \int e^{-m(\bar\lambda)/T} d\bar \lambda \ ,
\label{dfu}
\ee
where $Q(T,\bar x)$ is the partition function and $\bar\lambda$ represents the set of all microeconomic parameters. Here we have introduced the 
notation $\bar x = x_1,x_2,...,x_n$ to denote the possible set of {\it macroeconomic}  parameters which can appear after the integration over the entire
domain of definition of the microeconomic parameters $\bar\lambda$. 

Following the standard procedure of statistical thermodynamics \cite{huang}, we introduce the concept 
of mean value $\left\langle g \right\rangle $ for any function $g=g(\bar\lambda)$ as
\be
\left\langle g \right\rangle = \int g \rho \ d\bar\lambda = \frac{1}{Q(T,\bar x)} 
\int g\,e^{-m(\bar\lambda)/T} d\bar \lambda \ .
\ee
These are the main concepts which are needed in statistical thermodynamics for the investigation of a system which depends on the temperature $T$ 
and macroscopic variables $\bar x$. Consider, for instance, the mean value of the 
money $\left\langle m \right\rangle  = \int m \rho d\bar\lambda$ and 
let us compute the total differential $d\left\langle m \right\rangle$:
\be
d\left\langle m \right\rangle = \int \left( m d\rho + \rho dm \right)d\bar\lambda = \int m d\rho d\bar\lambda +\left\langle d m \right\rangle \ .
\ee
The first term of this expression can be further manipulated by using the definition of the density function (\ref{dfu}) in the form
$m=-T(\ln\rho + \ln Q)$. Then, we obtain (recall that $\int \rho d\bar\lambda = 1$ and, therefore, $d\int \rho d\bar\lambda =0$)
\be
d\left\langle m \right\rangle = TdS -  \sum_{i=1}^n y_i dx_i \ ,
\label{first}
\ee
where the entropy $S$ and the ``intensive" macroscopic variables are defined in the standard manner as 
\bea
S & = & \left\langle -\ln \rho  \right\rangle = \int (-\ln\rho) \rho \ d\bar\lambda\ ,\\
y_i &= &\left\langle - \frac{\partial m}{\partial x_i}  \right\rangle =  \int \left(- \frac{\partial m}{\partial x_i}\right) \,\rho \ d\bar\lambda\ .
\eea
Clearly, Eq.(\ref{first}) represents the first law of thermodynamics. Since the definition of temperature and entropy are in accordance 
with the concepts of statistical mechanics, the remaining laws of thermodynamics are also valid. 
Similar results  can be obtained for any quantity that can be
shown to be conserved in an economic system. This reflects the fact that different thermodynamic potentials can be used
to describe the same thermodynamic system. 

It is useful to calculate explicitly the entropy $S= \int (-\ln\rho) \rho \ d\bar\lambda$ by using the definition (\ref{dfu}) in the form 
$-\ln\rho = \ln Q + m/T$. The result can be cast in the form
\be
\label{freem}
f:= \left\langle m \right\rangle - TS = -T\ln Q(T,\bar x)  \ ,
\ee
so that
\be
S = - \frac{\partial f}{\partial T} , \quad y_i =  - \frac{\partial f}{\partial x_i} \ .
\label{thv}
\ee
This means that the entire information about the thermodynamic properties of the system is contained in the expression for  the 
``free money" $f$ which, in turn, is completely determined by the partition function $Q(T,\bar x)$. In statistical physics 
this procedure is still used, with excellent results, to investigate the properties of thermodynamic systems. We propose to 
use a similar approach in econophysics. In fact, to investigate a model for an economic system one only needs to formulate 
the explicit dependence of any conserved quantity, say money $m(\bar \lambda)$, in terms of the microeconomic parameters $\bar \lambda$. 
From $m(\bar \lambda)$ one calculates the partition function $Q(T,\bar x)$ and the free money $f(T,\bar x)$ which contains all
the thermodynamic information about the economic system.
  
In the next sections we present several examples of  hypothetical economic systems with relatively simple expressions for 
$m(\bar \lambda)$. In real economic systems, probably very complicated expressions for $m(\bar\lambda)$ will appear for which 
analytical computations are not available. Nevertheless, the calculation of the above integrals for the partition function 
can always be performed by using numerical methods so that the corresponding thermodynamic properties of the system 
can be found qualitatively.

\section{Hypothetical economic systems}
\label{sec:hyp}

The determination of the quantity $m$ in terms of the microeconomic parameters $\bar\lambda$ is a task that requires 
the knowledge of very specific conditions and relationships within a given economic system. The first step consists in  
identifying the microeconomic parameters $\bar\lambda$ which influences the capacity of an individual agent to 
compete for a share $m$ of the conserved quantity $M$. Then, it is necessary to establish how this influence should 
be represented mathematically so that $m$ becomes a well-defined function of $\bar\lambda$. 

Another aspect that must be
considered is the fact that in a realistic economic system not all agents are equivalent. For instance, an agent 
represented by an individual who works at a factory for a fixed yearly income would be considerably different from 
an agent represented by the firm to which the factory belongs. An important result of econophysics is that the 
Botzmann-Gibbs distribution is not affected by the specific characteristics of the agents involved in the economic model
\cite{yakov00}. For the 
statistical thermodynamic approach we are proposing in this work this means that we can decompose the quantity $m$ into 
classes $m=m^I + m^{II} + \cdots $, and different  classes can be described by different functions of different sets of microeconomic 
parameters. The formalism of statistical thermodynamics allows us to consider, in principle, all possible economic configurations
as far as $m$ is a well-defined function of $\bar\lambda$.

In the following subsections we will study several hypothetical economic systems in which $m$ is given as simple ordinary 
functions of the microeconomic parameters. 
We expect, however, that these simple examples will find some applications in the 
context of economic systems with sufficiently well defined microeconomic parameters. 
Although the function $m$ can represent any conserved economic quantity, for the 
sake of concreteness, we will assume that it represents the money and from now $m(\bar\lambda)$ will be referred as
to the {\it money function}. 

\subsection{Linear systems}
\label{sec:lin}

The simplest model corresponds to the case $m=c_0=$ const. Then, the partition function (\ref{dfu}) is given by
\be
Q(T,\bar\Lambda) = e^{-c_0/T} \bar\Lambda\ , \quad \bar\Lambda = \Lambda_1\Lambda_2 \cdots \Lambda_n \ ,
\label{pfconst}
\ee
where $\Lambda_i = \int d\lambda_i$, $i=1,2,\cdots ,n$, represent the macroeconomic parameters. The calculation of the thermodynamic 
variables, according to Eqs.(\ref{thv}), yields
\be
\label{thermconst}
f= c_0 - T\ln\bar\Lambda \ , \quad S = \ln \bar\Lambda\ ,\quad y_i = - \frac{\partial f}{\partial \Lambda_i} = \frac{T}{\Lambda_i} \ .
\ee
Furthermore, from the definition of free money $f$ we conclude that $\left\langle m \right\rangle = c_0$, i.e., the mean value of money
is a constant, as expected. This economic model is considerably simple. Each agent possesses the same amount of money $c_0$, the entropy does not
depend on the mean value of the money $c_0$, and the state equations are $\left\langle m \right\rangle = c_0$ and $y_i \Lambda_i = T$. The system 
is completely homogeneous in the sense that each agent starts with a given amount of money, $c_0$, and  ends up with the same amount. 
Probably, the only way to simulate such an economic system would be by demanding that agents do not interchange money; this is 
not a very realistic 
situation. Indeed, the fact that the entropy is a constant, that does not depend on the mean value of the money, allows us to renormalize the macroscopic
parameters in such a way that $\Lambda_i = \int d\lambda_i = 1$, for each $i$, so that the total entropy vanishes. In this case, from 
Eq.(\ref{thermconst}) we see that $f=c_0$ and the corresponding equations of state are compatible with the limit $T\rightarrow 0$.
This resembles the argumentation used in the description of the third law of thermodynamics. This observation indicates that 
a completely homogeneous economic system is not realizable as a consequence of the third law of thermodynamics.    

Consider now the function 
\be 
m=c_1\lambda_1\ ,
\ee
where $c_1$ is a positive constant. The corresponding partition function can be written as
\be
\label{pflin1}
Q(T,\bar \Lambda) = \int e^{-c_1\lambda_1/T}d\bar\lambda = \frac{\bar\Lambda}{\Lambda_1}\int_0^\infty e^{-c_1\lambda_1/T}d\lambda_1 = 
\frac{T\bar\Lambda}{c_1\Lambda_1} \ .
\ee
The relevant thermodynamic variables follow from Eqs.(\ref{freem}) and (\ref{thv}) as
\bea
f & = & -T \ln\frac{T\bar\Lambda}{c_1\Lambda_1}  \ ,\quad
S = 1 +  \ln \frac{T\bar\Lambda}{c_1\Lambda_1}  \ ,\\   
y_1 & =& 0\ , \quad y_j= \frac{T}{\Lambda_j}\ , \ j\neq 1\ ,
\label{thvlin}
\eea
and the conservation law (\ref{first}) becomes
\be
d\left\langle m \right\rangle = T d S - T\sum_{j=2}^n \frac{d\Lambda_j}{\Lambda_j}
\ee
Moreover, comparing the above results with the definition of $f$, it can be shown that  
$\left\langle m \right\rangle = T$, and  so the fundamental thermodynamic equation in the entropic representation
\cite{callen} can be written as
\be
\label{fel1}
S = 
1 + \ln\frac{\left\langle m \right\rangle  }{c_1} + \sum_{j=2}^n\ln\Lambda_j \ .
\ee
This expression relates all the extensive variables of the system and it can be used to derive all the equations of state 
in a manner equivalent to that given in Eq.(\ref{thvlin}). Notice that in this case the entropy is proportional to the
temperature (mean value of money) so that an increase of the average money per agent is necessarily associated with an 
increase of entropy. This observation is in agreement with the second law of thermodynamics. Notice, furthermore, that 
in a limiting case, similar to that considered in the first example given above, 
it is possible to renormalize the macroeconomic parameters 
$\Lambda_j$ such that the last term of the fundamental equation (\ref{fel1}) vanishes. Nevertheless, in order to  
reach the minimum value of the entropy it is necessary to consider the limit $T\rightarrow 0$. Again, we consider this 
result as an indication of the validity of the third law of thermodynamics.

In Eq.(\ref{pflin1}) we have chosen the interval $[0,\infty)$ for the integration along the  variable $\lambda_1$. 
As a consequence the macroscopic variable $\Lambda_1$ vanishes from the final expression for the fundamental 
equation (\ref{fel1}), and consequently $y_1=0$. 
 However, it is also possible to consider the interval $[\lambda_1^{min},\lambda_1^{max}]$
so that the macroscopic variables $\lambda_1^{min}$ and $\lambda_1^{max}$  reappear in the fundamental equation and
can be used as extensive variables which enter the conservation law (\ref{first}). In a realistic economic system 
the interval of integration will depend on the economic significance of the microeconomic parameter $\lambda_1$. 
For the sake of simplicity, we choose in this work the former case in which the corresponding macroeconomic parameter
does not enter the analysis.

It is interesting to analyze the most general linear system for which 
\be
m = c_0 + \sum_{i=1}^n c_i \lambda_i \ ,
\ee
where $c_0, c_1, ...$ are positive real constants.  It is then straightforward to calculate the partition function
\be
Q(T) = \frac{1}{\bar c} e^{-c_0/T} T^n \ , \qquad \bar c = c_1 c_2 \cdots c_n \ ,
\ee
and the relevant thermodynamic variables
\bea 
f&=& c_0 -T \ln \frac{T^n}{\bar c} \ , \quad S = n+ \ln \frac{T^n}{\bar c}\ , \\ 
y_i &=&0 \ ,\quad
 \left\langle m \right\rangle = c_0 +n T\ .
\eea
All the macroscopic parameters vanish and the system depends only on the temperature. However, the total number of macroscopic 
parameters $n$ does enter the expression for the entropy so that to increase the mean value of the money by the amount 
$\Delta \left\langle m \right\rangle = n(T_2 - T_1)$, it is necessary to increase entropy by an 
amount $\Delta S = n \ln(T_2/T_1)$; both amounts are proportional to the total number of macroscopic parameters. 

Another 
consequence of this analysis is that once the constants $c_0$ and $n$ are fixed,
 it is not possible to change the mean value of the money 
without changing the temperature of the system. This means that  an isothermal positive change of  $\left\langle m \right\rangle$
is possible only by increasing the total amount of money in the system. 

\subsection{Nonlinear systems}
\label{sec:nonl}

Consider the quadratic function $m=c_1\lambda_1^2$ which generates the partition function 
\be
Q(T,\bar \Lambda)= \left(\frac{\pi}{c_1}\right)^{1/2} \frac{\bar\Lambda}{\Lambda_1} T^{1/2} \ ,
\ee
where we have considered the microeconomic parameter $\lambda_1$ in the interval $(-\infty,\infty)$. 
The corresponding thermodynamic variables are 
\be
S=\frac{1}{2}\left( 1+ \ln \frac{\pi T}{c_1}\right) + \ln\frac{\bar\Lambda}{\Lambda_1} \ ,\ y_1=0\ ,\ 
y_j=\frac{T}{\Lambda_j}\  ,\ \left\langle m \right\rangle =\frac{T}{2}\ ,
\ee
which can be put together in the fundamental equation
\be
S=\frac{1}{2}\left( 1+ \ln \frac{2\pi \left\langle m \right\rangle}{c_1}\right) + \ln\frac{\bar\Lambda}{\Lambda_1} \ .
\ee
Again we see that the effect of considering the extreme values of the parameter $\lambda_1$ is that the 
corresponding macroscopic variable $\Lambda_1$ does not enter the expressions for the thermodynamic variables and, consequently,
the corresponding intensive thermodynamic variable vanishes. Furthermore, it is evident that 
the power of $\lambda_1$ in the money function leads to a decrease of the mean value $\left\langle m \right\rangle$. 

To investigate the general case,
we consider the monomial functional dependence $m = c_1\lambda_1^{\delta}$, with $\delta$ being an arbitrary real constant. 
A straightforward calculation leads
to the following partition function and thermodynamic variables:
\be
\label{pfnl2}
Q(T,\bar\Lambda)  =  \frac{\bar\Lambda}{\delta \Lambda_1} \left(\frac{T}{c_1}\right)^{1/\delta} \Gamma\left(\frac{1}{\delta}\right)
 \ , 
\ee
\be
S  = \frac{1}{\delta}\left(1+\ln \frac{T}{c_1}\right) + \ln \left[\frac{\bar\Lambda}{\delta\Lambda_1} \Gamma\left(\frac{1}{\delta}\right)\right]
 \ ,\ 
\left\langle m \right\rangle =\frac{T}{\delta}\ ,
\label{tvlin2}
\ee
and the intensive variables $y_i$ are given as in the previous case. 
It follows that $\delta >0$ in order for the mean value of the money to be positive. If $\delta>1$, the mean value of the money decreases, whereas 
it increases for $0<\delta<1$. In such a hypothetical system, a way to increase the amount of money per agent would be to identify the 
microeconomic parameter $\lambda_1$ and apply the measures which are necessary for the money function $m$ to become $m\propto \lambda_1^\delta$,
with $\delta<1$.  

If we consider a transition of an economic system from a state characterized by the parameter $\delta_1$ to a new state with parameter
$\delta_2$, maintaining the same temperature, the mean value of the money undergoes a change $\Delta \left\langle m \right\rangle = 
(1/\delta_2-1/\delta_1)T$, so that for a positive change we must require that $\delta_2<\delta_1$. Moreover, if we desire a positive change 
by an amount greater than the average money per agent ($\Delta \left\langle m \right\rangle > T$), we must demand that 
$\delta_2 < \delta_1/(1+\delta_1)$.  Even if 
the initial state corresponds to a linear system $(\delta_1 = 1)$, in which no increase of  $\left\langle m \right\rangle$ is possible, 
we can reach a state of greater  mean value $\left\langle m \right\rangle$ by demanding that $\delta_2< 1/2$.  Of course, 
for a positive change of 
$\left\langle m \right\rangle$ the ``price" to be payed  will result in an increase of entropy by an amount which is proportional to
the coefficient $(1/\delta_2-1/\delta_1)$.

We see this possibility as an advantage of our
statistical approach. In fact, we start from an equilibrium state with a lower value of $\left\langle m \right\rangle$ and end up in a state
with a higher value of $\left\langle m \right\rangle$, by choosing appropriately a particular microeconomic parameter. The positive change
of the macroeconomic variable $\left\langle m \right\rangle$ is induced by a change of a microeconomic parameter. Once the process is started,
the system will naturally evolve into a state characterized by a higher value of $\left\langle m \right\rangle$. This natural evolution
occurs because, as has been repeatedly shown in computer simulations \cite{yakov07}, 
the final equilibrium state corresponds to a Boltzmann-Gibbs probability distribution, 
which is a basic component of the approach proposed in this work.

\subsection{Multiplicative systems}
\label{sec:mul}

An interesting case follows from the money function $m(\bar\lambda) = c_1 \ln \lambda_1$. In fact, the partition function is given by
\be
Q(T,\bar \Lambda) = \frac{\bar\Lambda}{\Lambda_1} \frac{1}{\alpha x ^{\alpha}}\ , \quad \alpha= \frac{c_1}{T} - 1>0  \ ,
\ee
where we have used $\lambda_1\in [x,\infty)$, with $x=\lambda_1^{min}$. This expression is nothing more but the partition function (cumulative probability) of the Pareto distribution density  $\rho_{_P} \propto 1/x^{1+\alpha}$ which has been shown to correctly 
describe the distribution of money (and other conserved economic quantities) in the upper tail of the distribution, i.e., 
for amounts greater than $x$. A similar derivation of the Pareto law was recently performed in \cite{pareto}.

We now have the possibility to analyze the Pareto distribution in terms of macroeconomic parameters. 
The computation of the thermodynamic variables using Eq.(\ref{thv}) yields 
\be
\label{entp}
S= \frac{c_1}{c_1-T} + \ln\frac{xT}{c_1-T} + \ln\frac{\bar\Lambda}{\Lambda_1}\ ,
\ee
\be
y = -\frac{\partial f}{\partial x} = -\frac{c_1-T}{x}\ , \ y_j = \frac{T}{\Lambda_j}\ .
\ee
Notice that the intensive variable $y$ conjugate to the lower limit $x$ of Pareto's distribution appears with a different sign,
when compared with the remaining intensive variables $y_j, \ j=2,3,...,n$. As a consequence of this change of sign 
the conservation law is given as
\be
d\left\langle m \right\rangle = T dS + \frac{c_1-T}{x} dx -  \frac{T}{\Lambda_j}d\Lambda_j\ ,
\ee
so that if we interpret, by analogy with physical systems, the intensive variables as ``forces", we can conclude that
the  ```force" corresponding to Pareto's distribution is negative. Moreover, if from the above expressions and that of the
free money $f$, we calculate the  mean value $\left\langle m \right\rangle$, we obtain
\be
\left\langle m \right\rangle =  \frac{c_1T}{c_1-T}+ c_1\ln x \ .
\ee
The origin of the second term is clear because we have chosen the money function as $m=c_1\ln \lambda_1$, and $x$ is the minimum value
of $\lambda_1$. The first term, however, is new and has the interesting property that it diverges as $c_1\rightarrow T$. This  means
that it is possible to increase arbitrarily the value of $\left\langle m \right\rangle$ in the upper tail of the distribution of money, 
 maintaining  the  values of $x$ and $T$ fixed, solely by fine tuning the value of the constant $c_1$ so that it takes the value of the average
amount of money per agent $T$. Perhaps this simple observation could be useful for the understanding of the monetary evolution that 
takes place in the upper class of the wealth distribution in certain economies. 

Finally, we analyze the case of the money function 
\be
m= c_1\lambda_1^\delta + d_1 \ln \lambda_1 \ ,
\ee
which corresponds to the density 
distribution
\be
\rho(m) = \lambda^\gamma e^{-\beta\lambda^\delta} \ ,  \quad \beta= \frac{c_1}{T} \ ,\quad \gamma = \frac{d_1}{T}\ ,
\ee
where $c_1$, $d_1$ and $\delta$ are constants. This expression is known in the literature as the Gamma distribution and has been
used in econophysics to investigate models with multiplicative asset exchange \cite{ikr98}. A straightforward computation shows that 
the corresponding partition function can be cast in the form
\be
\label{pfgamma}
Q(T,\bar\Lambda)  =  \frac{\bar\Lambda}{\delta \Lambda_1} \left(\frac{T}{c_1}\right)^{(1-d_1/T)/\delta} 
\Gamma\left[\frac{1}{\delta}\left(1-\frac{d_1}{T}\right)\right]
 \ , 
\ee
an expression which is essentially equivalent to the partition function (\ref{pfnl2}) following from the monomial function 
$m\propto \lambda_1^\delta$ discussed in Sec. \ref{sec:nonl}. In particular, in the limit $d_1<<T$ the two partition functions coincide. 
Strong differences can appear near the points where the function (\ref{pfgamma}) diverges. This occurs when $T\rightarrow 0$, which we should 
avoid in accordance with the third law of thermodynamics,  and at the poles of
the Gamma function, i.e, for $(1-d_1/T)/\delta=-k=0,-1,-2,\cdots$ The last possibility, however, is not necessarily in disagreement with the
third law of thermodynamics since at the poles the temperature can be made to tend to a constant positive value, $T\rightarrow d_1/(1+k\delta)$. 

It is not difficult to derive the thermodynamic variables  for this case. The singular pole structure of the partition function leads to 
a peculiar behavior of the thermodynamic variables which deserve a more detailed and deeper analysis. Outside the poles, however, the 
thermodynamic behavior is essentially dictated by  Eqs.(\ref{tvlin2}) which correspond to a Boltzmann-Gibbs distribution with a power law
dependence for the money function. This result shows that from a macroeconomic point of view there is no essential difference between 
the Gamma law distribution and the Boltzmann-Gibbs distribution, as far as the value for the temperature does not correspond to a 
 singular pole of the partition function of the Gamma distribution. This result explains in a simple manner why in concrete examples
it is possible to mimic the results of an exponential distribution by choosing appropriately the additional parameters of the Gamma law
distribution \cite{yakov06}.

\section{Phase transitions}
\label{sec:ptr}

An important feature of many thermodynamic systems is their capability to exist in different phases with specific interior and exterior
characteristics which can drastically change during a phase transition. The thermodynamic approach proposed in this work insinuates 
the possibility of considering the phase structure of  economic systems and the conditions under which a particular economic system can undergo a
phase transition. In this section we will perform such an analysis for the hypothetical systems studied above.

Recall that a phase transition is usually associated with discontinuities or divergencies in the thermodynamic variables or its derivatives. 
In particular, the behavior of the entropy function is used as a  criterion for the analysis of phase transitions. Moreover, the heat capacity 
\be
C=T\,\frac{\partial S}{\partial T}
\ee
is an important thermodynamic variable which indicates the existence of second-order phase transitions. 

From the results presented in the preceding sections one can show that the heat capacity 
for systems characterized by a monomial money function, $m=c_1\lambda^\delta$,
is constant, namely $C=1/\delta$, with $C=0$ for the limiting case $m=c_0= $ const. Since the constant $\delta$ must be positive for the
 mean value $\left\langle m \right\rangle$ to be positive, the temperature of such systems raises 
 under an increase of ``economic heat", and vice versa. 
An inspection of the remaining thermodynamic variables shows that such systems cannot undergo a phase transition. 
An interesting additional result is that such hypothetical systems are stable. In fact, a positive heat capacity is usually interpreted 
in statistical mechanics as a condition of stability. Consequently, systems described by monomial money functions with $\delta>0$ are stable. 

The situation is 
different in the case of multiplicative systems. For a system characterized by a Pareto law distribution, we obtain from Eq.(\ref{entp})
\be
C_x = \left(\frac{c_1}{c_1-T}\right)^2 \ .
\ee
First, we notice that this heat capacity is positive, indicating that the system is stable. However, 
a phase transition occurs in the limit $c_1\rightarrow T$ where the heat capacity diverges.
This is also the value for which the mean value of the money raises unlimitedly. 
This shows that a phase transition can be induced by choosing appropriately the parameter $c_1$ and the corresponding economic system
undergoes a drastic change with agents possessing more and more money; however, this hypothetical process must end at some stage due 
to the natural boundary imposed by the fact that the total amount of money in the system is fixed and finite. 
During the phase transition several thermodynamic variables diverge and, consequently,  
the thermodynamic approach breaks down. As in ordinary thermodynamics, a different (nonequilibrium) approach is
necessary to understand the details of the phase transition. This is beyond the scope of the present work.

In the case of a system with a Gamma law distribution and partition function given in Eq.(\ref{pfgamma}), the phase structure is much more 
complex. The analytical results are rather cumbersome expressions that cannot be written 
in a compact form. A preliminary numerical analysis shows 
that  near the poles of the partition function first-order and second-order phase transitions 
occur with two possible scenarios. The first one is similar to the phase transition of a system with a Pareto law distribution with 
the  mean value $\left\langle m \right\rangle$ increasing either exponentially or as $\left\langle m \right\rangle\propto 1/(d_1-T)^c$, and $d_1\rightarrow T$,
 where the value of 
the constant $c$ depends on the kind of pole of the partition function. The second scenario is characterized by a rapid reduction of 
 $\left\langle m \right\rangle$, tending exponentially to a fixed positive value $m_0$ 
 which depends on the kind of pole. Unexpectedly, we also found divergencies in the corresponding heat capacity which are not related
to the poles of the partition function. A more deep analysis will be necessary to understand the complete phase structure of 
this specific  multiplicative system. 

\section{Discussion and conclusions}
\label{sec:con}

In this work we propose to apply the standard methods of statistical thermodynamics in order to investigate the structure and behavior 
of economic systems. The starting point is an economic system in which a conserved quantity is present. In certain current economies it 
has been shown that such quantities exist, the money being one of them. We have shown that to any  conserved quantity it is possible 
to associate a function $m$ (the money function, for instance) 
 from which all the thermodynamic variables and properties of the system can be derived.
The money function depends on a set of microeconomic parameters which generate macroeconomic parameters at the level of the partition 
function. The thermodynamic variables depend on the macroeconomic parameters and satisfy the ordinary laws of thermodynamics. 

Starting from simple money functions, 
we consider linear, nonlinear and multiplicative systems as examples of hypothetical economic systems in which it is possible to 
apply our approach, obtaining analytical results. In all the cases, we computed the most relevant thermodynamic variables and 
analyzed their behavior. The results show that it is possible to manipulate the microeconomic parameters in order to control 
the output at the level of the macroeconomic parameters.  
We see this possibility as an advantage of our
statistical approach. One can start from an equilibrium state with some given  mean value for the money $\left\langle m \right\rangle$,
 and raise its
value by choosing appropriately a particular microeconomic parameter. Once the process is started,
the system will naturally evolve into a state characterized by a greater value of $\left\langle m \right\rangle$. 
This natural evolution occurs because, as has been repeatedly shown in computer simulations \cite{yakov07}, 
the final equilibrium state corresponds to a Boltzmann-Gibbs probability distribution, 
which is a basic component of the approach proposed in this work. This evolution, however, must be understood as 
in standard thermodynamics, i.e., the evolution process must be quasi-static so that at each step the economic system
is in equilibrium. A more realistic evolutionary model must take into account nonequilibrium states, a task that cannot
be treated within the standard approach of statistical thermodynamics. It would be interesting to investigate if 
the existing generalizations of nonequilibrium thermodynamics can also be applied  in the context of economic systems.

We propose to use the formalism of phase transitions to analyze the behavior of economic systems. The relatively 
simple examples studied in this work show that in fact a phase transition can  be associated to drastic changes of 
the  mean value of the money. Economic crisis are usually accompanied by far-reaching modifications of some macroeconomic 
variables. It would be interesting to propose more sophisticated and realistic models for money functions and 
investigate their behavior during phase transitions. If a crisis could be understood this way, an appealing problem would 
be to explore the possibility of controlling its consequences. To investigate this problem it will be necessary to 
explore and establish the economic meaning of the microeconomic variables $\lambda_l$ by using an approach based upon
concepts of standard economics.  This remains an open question that could be the subject of further investigations. 

In the context of systems with many constituents, thermodynamic interaction is an important concept which can 
 completely modify the interior and exterior structure of the system. 
 In physics, one can introduce thermodynamic interaction into a system
by choosing appropriate potentials, because we understand the physical meaning of the Hamiltonian function. 
In our approach the analogous of the Hamiltonian is the money function; we believe, however, that its economic 
significance, at least at the moment,  is much more sophisticated. Therefore, 
we propose to use a different method to introduce thermodynamic interaction
into an economic system. Recently, the theory of geometrothermodynamics \cite{quev07} 
was formulated with the aim of describing  thermodynamics in terms of geometric concepts. One of the results of this
formalism is that thermodynamic interaction can be interpreted as the curvature of the equilibrium space. This has 
been shown to be true not only in the case of ordinary thermodynamic systems, like the ideal gas and the van der Waals gas \cite{qsv08,vqs08},
but also in more exotic configurations like black holes \cite{aqs08}. This opens the possibility of introducing
thermodynamic interaction by just modifying the curvature of the equilibrium space. In fact, it can be shown that all 
the hypothetical economic systems analyzed in this work have very simple equilibrium spaces for which the manipulation 
of curvature is not a difficult task. We expect to analyze this possibility in the near future.

\section*{Acknowledgements} 
This work was supported in part by DGAPA-UNAM, grant No. IN106110. One of us (HQ) would like to thank 
G. Camillis for interesting comments and suggestions.


\end{document}